A STUDY OF SOLAR PHOTOSPHERIC TEMPERATURE GRADIENT

VARIATION USING LIMB DARKENING MEASUREMENTS

Serena Criscuoli, National Solar Observatory, Boulder, Colorado, 80303

and

Peter Foukal, Nahant, Massachusetts, 01908

ABSTRACT

The variation in area of quiet magnetic network measured over the sunspot cycle should modulate the spatially averaged photospheric temperature gradient, since temperature declines with optical depth more gradually in magnetic flux tube atmospheres. Yet, limb darkening measurements show no dependence upon activity level, even at an rms precision of 0.04%. We study the sensitivity of limb darkening to changes in area filling factor using a 3-D MHD model of the magnetized photosphere. The limb darkening change expected from the measured 11- yr area variation lies below the level of measured limb darkening variations, for a reasonable range of magnetic flux in quiet network and internetwork regions. So the remarkably constant limb darkening observed over the solar activity cycle is not inconsistent with the measured 11 – year change in area of quiet magnetic network. Our findings offer an independent constraint on photospheric temperature gradient changes reported from measurements of the solar spectral irradiance from the Spectral Irradiance Monitor (SIM), and recently, from wavelength – differential spectrophotometry using the Solar Optical Telescope (SOT) aboard the HINODE spacecraft.

1.INTRODUCTION

The decrease of photospheric brightness towards the solar limb led to the finding that solar energy is transported to the sun's visible surface mainly by radiation (Schwarzschild 1906). Since then, increasingly accurate measurements of the wavelength dependence of this center to limb variation (CLV) have provided basic information on photospheric structure and chemical composition (e.g. Minnaert 1953).

More recently, interest has turned to the information that time variation in the limb darkening might provide on changes in total and spectral solar irradiance (Petro, Foukal & Kurucz 1985). For instance, Petro et al. (1984) used the constancy of the CLV they measured in 1980 – 82 to pose a constraint on an unusually large change in UV spectral irradiance suggested by published measurements of Fraunhofer line equivalent widths over that period. Several sets of measurements since 1974 have shown that the limb darkening is surprisingly constant over a wide range of solar activity (Pierce & Slaughter 1977; Petro et al. 1984; Neckel & Labs 1994; Livingston & Wallace 2003; Elste & Gilliam 2007; Livingston & Milkey 2009). Several error sources might be responsible for a spurious limb darkening variation (Petro et al 1984; Neckel & Labs 1994; Foukal 1989) but a null result is harder to dismiss, especially since it is found by independent observers using different apparatus and procedures over three sunspot cycles.

This constancy of the limb darkening took on renewed interest with the report by Harder et al. (2009) of a solar dimming with increased activity at wavelengths formed deep in the photosphere, together with a brightening at those originating at higher levels. Harder et al. attributed this decreased photospheric temperature gradient to the increase in facular area filling factor with increasing activity, since temperature is known to decrease more slowly with optical depth in the facular atmosphere than in the quiet photosphere (e.g. Chapman 1979; Spruit 1976). Alternatively, they suggested that it might represent a global change in photospheric thermal structure due to variations in mechanical energy transport by magnetic free energy or convective motions. These explanations have since been explored in more detail by Fontenla et al. (2011), Criscuoli & Uitenbroek (2014a) and by Fontenla et al. (2015).

Harder et al. based their study on data from the Spectral Irradiance Monitor (SIM) spectrometer carried on the Solar Radiation and Climate Experiment (SORCE) (Rottman, 2005). The calibration of the SIM has proven difficult (e.g. Béland et al. 2014, Ermolli et al. 2013) so their finding remains controversial. But their point that the photospheric temperature gradient is modulated by the changing area of magnetic flux tubes over the sunspot cycle must hold, at some amplitude level.

In Section 2 we estimate the gradient change that might be expected from measurements of the lower facular temperature gradient and the change in quiet network area between a minimum and maximum of solar activity. In Section 3 we discuss the measurements of limb darkening since 1974. We describe our calculations of the sensitivity of limb darkening to network area in Section 4. We discuss our results in Section 5 and present our conclusions in Section 6.

2. 11- YR MODULATION OF THE PHOTOSPHERIC TEMPERATURE GRADIENT EXPECTED FROM CHANGE OF FACULAR AREA.

Semi- empirical modeling of facular atmospheres (e.g. Chapman 1979; Vernazza, Avrett & Loeser 1981; Fontenla et al., 2006) indicates that the temperature gradient of the facular atmosphere over the depth range responsible for the visible and near IR spectral range is about 20% lower than in the surrounding photosphere. A similarly lower gradient is measured directly from differential photometry of faculae in continuum (Foukal and Duvall 1985) and in Fraunhofer lines (Elste 1985). Such a lower gradient is also expected from static and dynamical models of flux tubes (e.g. Spruit 1976; Steiner et al. 1998; Steiner 2005; Criscuoli & Rast 2009) and from magneto-hydrodynamic simulations of the solar photosphere (e.g. Carlsson et al. 2004, Criscuoli 2013).

The area filling factor of the quiet network measured from Kitt Peak magnetograms increased from about 15% to 20% from the minimum to the maximum of cycle 21 (Foukal, Harvey & Hill 1991). This 5% increase, together with the 20% lower gradient in the network flux tubes, suggest a globally averaged decrease of about 1% of the temperature gradient over a sunspot cycle, due to area change of the quiet network alone.

About half of this area and gradient change would be expected over the 2004 - 2007 period when the Harder et al. data were obtained. The range in brightness temperature over the photospheric depth interval studied by Harder et al. is approximately 1000 K (e.g. Vernazza, Avrett and Loeser 1981), so the slightly less than 10 K change they report in Fontenla et al. (2011) represents a fractional gradient change of order 1% during those 3 years of the cycle 23 decline. A temperature gradient change of the sign and magnitude reported by Harder et al. could, therefore, be produced by the measured lower temperature gradient of facular flux tubes and the measured 11 year change in quiet network area alone.

But approximately half of the 11 year variation of full disk chromospheric indices like F10.7 or disk – integrated Ca K is caused by changing area of active region plage and half by the combined area change of the quiet and active network (e.g. Foukal & Lean 1988). It follows that the change of quiet network filling factor measured by Foukal, Harvey & Hill (1991) represents less than ½ of the *total* area change of facular magnetic flux tubes. This suggests that the 11- year change in total area of magnetic flux tubes could easily account for a temperature gradient change even larger than reported by Harder et al.

## 3. CONSTRAINTS FROM PHOTOSPHERIC LIMB DARKENING MEASUREMENTS

Precise limb darkening measurements have been made since 1974 using the McMath telescope and spectrograph at Kitt Peak (Pierce and Slaughter 1977; Petro et al. 1984; Neckel and Labs 1994; Livingston & Wallace 2003; Livingston & Milkey 2009). A separate set of observations was obtained at the Sacramento Peak coronagraph, to minimize scattering (Elste and Gilliam 2007). The distribution in time and activity level of observations at the 445.1 nm continuum wavelength shared by these observation sets, is shown in Table 1.

Searches for cycle dependence of these limb darkening curves, described in these studies, have failed to produce any positive detections. The measurements by Petro et al. between late 1980 and late 1982 span about half of the activity range of cycle 21. Those of Elste and Gilliam in 1988 and 1990 cover a somewhat greater activity range in cycle 22. The Neckel and Labs data include measurements made at the 1986 minimum and the 1990 maximum of cycle 22, so they span the largest activity range. These three data sets have been corrected for scattered light and averaged over several scans to reduce granulation and 5 – minute oscillation noise, so they are the most suitable for study of long term variations.

The remarkably small variation of limb darkening measured in these observations is shown in Fig 1. The rms relative difference between the 1988 and 1990 coronagraph data is less than 0.05%. The same is true of the internal scatter in the 1980 through 1982 McMath data of Petro et al. Neckel and Labs (1994) report no change between 1986 and 1990 exceeding 0.04%. Most impressive, though, is the *absolute* agreement to 0.04% rms between the limb darkening curves measured a decade apart by Petro et al. and Elste and Gilliam; independent observers using completely different instruments and procedures. (See Elste and Gilliam 2007).

The precision and reproducibility of limb darkening measurements are affected by several error sources (e.g. Petro et al.1984; Foukal 1989; Neckel and Labs 1994). Granulation and p- mode noise can be reduced by averaging many scans. Atmospheric and instrumental blurring can be corrected to various degrees depending on the observing technique; scattering can be neglected in coronagraph observation at high altitude. Detector hysteresis, spectral passband stability and fluctuations in scan speed (caused by e.g. wind loading on the McMath heliostat mirror) must also be considered. Any inclusion of active network or active region faculae in the scans would generate spurious variations in limb darkening, so they also do not help explain the absence of variation.

There are, therefore, many possible sources of small variations. However, the result presented here is the *absence* of any change. Such a null result is harder to explain as a measurement error.

4. THE SENSITIVITY OF LIMB DARKENING TO FLUX TUBE AREA FILLING FACTOR

The sensitivity of limb darkening to a change in photospheric temperature gradient was studied by Petro, Foukal & Kurucz (1985). The sensitivity found using one and two- component radiative- convective models, was approximately 0.7% change in limb darkening at 445.1 nm for a 10 K change in temperature gradient around $\tau_{0.5} = 1$. This is an order of magnitude larger than the precision of the limb darkening measurements described above, and should be easily detectable. But 3-D MHD modeling has since shown (Uitenbroek & Criscuoli 2011) that the dependence is complicated by granule geometry and cannot be calculated reliably from such semi-empirical models.

In the present study we have used the results from runs of the Copenhagen – Stagger 3-D MHD code (Galsgaard & Nordlund 1996) to investigate the differences in limb darkening produced by changes in flux tube filling factor consistent with the measurements described above in Section 3. Following the procedure described previously by Criscuoli (2013), we considered a set of 10 hydrodynamic (HD, thereafter) snapshots and two sets of 10 MHD snapshots initialized with vertical unipolar magnetic intensity values of about 50 and 100 G, respectively. As described in previous studies (e.g. Schüssler and Vögler 2008, Nordlund et al. 2009), the initial vertical magnetic field is advected toward intergranular lanes, where it forms "sheet-like" and "micropore" kG structures, surrounded by weak or relatively field-free regions. In this reorganized state the field is still mostly vertical and the values of the spatial average of the magnetic field intensity over the optical depth unity surface is close to the initial one (Fabbian et al. 2012). The snapshots simulated a 6 x 6 $Mm^2$ portion of the photosphere, horizontally sampled at 24 km /pixel; the vertical resolution was not constant but equal to about 15 km/pixel at the base of the photosphere. Each snapshot had original dimensions of 252 x 252 pixels horizontally and 126 vertically.

For our analysis we considered a portion of the original snapshots along the vertical direction which included 500 km above and below (for a total of 1 Mm) the height that corresponded to an average optical depth of unity at 500 nm in the HD snapshots. The snapshots are described in detail in Fabbian et al. (2010, 2012) and have been employed to investigate the relevance of magnetic fields in abundance determinations (Fabbian et al., 2010, 2012), temperature fluctuations in quiet photospheric regions (Beck et al., 2013), properties of G-band bright points (Criscuoli & Uitenbroek, 2014b), plasma properties in quiet and active regions (Criscuoli, 2013), and to interpret SIM measurements at visible and infrared bands (Criscuoli & Uitenbroek 2014a). We then employed the RH code (Uitenbroek 2001) to synthesize the emergent continuum intensity radiation at 445.125 nm for ten different lines of sight in each of the snapshots.

As discussed in Criscuoli & Uitenbroek (2014a) and Criscuoli (2013) the average temperature gradients in the simulations decreased with the increase of the average magnetic flux. Similarly, the temperature gradient, $dT/d\tau$, in the magnetic features found in both the 50 G and 100 G snapshots was approximately 50% lower than in the non- magnetic (HD) photosphere, over the range of 500 nm optical depth $-1.0 < \text{Log } \tau < 1.0$ covered by the limb darkening observations (Figure 2). This is consistent with the lower gradient found from the observations and modeling described in Section 2 above. The small differences between the temperature gradients found for magnetic structures singled out on different magnetic flux simulations results from the different level of suppression of convection, as discussed in detail in Criscuoli (2013).

Simulations obtained with similar magnetic field intensities and configurations have already been employed to investigate properties of quiet Sun regions – internetwork and quiet network (e.g. Hirzberger et al. 2010, Afram et al 2011, Beck et al. 2013, Riethmüller et al. 2013). In a recent paper, Fabbian & Moreno-Insertis (2015) showed that intensity CLV in continua obtained with the same set of (M)HD simulations employed in this study agrees with observations reported in Neckel & Labs (1994) and Pierce et al. (1977). They reported no significant difference at the 1% level between limb darkening curves obtained from snapshots characterized by different amount**s** of magnetic flux.

Fig. 3 shows that the differences between the continuum intensity CLV computed from the different snapshots are indeed below 0.015, after normalization for the disk center intensity. But these differences are still an order of magnitude larger than those observed between high and low solar activity. So it appears that the various HD and MHD snapshots represent magnetic regimes that are much more widely separated than the quiet sun regimes observed around minima and maxima of the 11 yr activity cycle. This appears, in itself, to be an interesting finding since it is not quite clear from existing discussions (e.g. Fabbian et al. 2012, Fabbian and Moreno Insertis 2015) whether the snapshots are to be interpreted as various levels of magnetization of the quiet sun, or of the quiet and active network.

To reproduce the small measured CLV differences we found it necessary to use the MHD snapshots to represent not the quiet sun, but rather the network and internetwork components of the quiet sun, weighted according to their observed area filling factors at activity minimum and maximum. The flux values that best represent the network and internetwork are still uncertain (Sánchez Almeida & Martínez González, 2011; Martínez Pillet, 2013). So we calculated the limb darkening separately using three separate assumptions: a) network: 100 G, internetwork: HD; b) network: 100 G, internetwork: 50 G; c) network: 50 G, internetwork: HD. (N.B.: These magnetic intensities refer to the values spatially averaged over the snapshot; the intensities in the flux tubes are of order 1kG, as observed in magnetograms at high angular resolution). The models are summarized in Table 2.

5. RESULTS AND ERRORS

The dependences of limb darkening upon the network area for the various models are shown in Figure 4, for comparison with the measurements. The plot shows the absolute variation of the intensity CLV (normalized to the disk center intensity) between the maximum and the minimum. The differences peak at $\mu \approx 0.3$, at values below 0.0007. Comparison with the measurements indicates that all the modelled CLV differences lie within the values observed by Elste & Gilliam and Petro et al.

These modeled CLV differences were calculated using a filling factor increase from 15% to 20% as measured over the full activity range in cycle 21 by Foukal, Harvey & Hill (1991). But the appropriate activity (and filling factor) range for the Elste & Gilliam and Petro et al. measurements is only about half of that. So the calculated CLV variation lies even farther below the measured variation than inferred from Figure 4. The variation calculated for model 1 lies above the upper limit set by Neckel & Labs in their measurements that spanned the full range of activity in cycle 22. This suggests that models 2 and 3, in which the difference in magnetic intensity is approximately 50 G are to be preferred over model 1 in which the magnetic intensity difference is about twice as large.

The Petro et al. measurements were carried out in a 0.02 nm continuum window centered at 445.125 nm with a spectral purity of between 0.005 and 0.015 nm. Calculations we performed indicate that accidental inclusion of the wing of a nearby Ti I line in the measurements should have minimal influence on the limb darkening. Also, any such line contamination would tend to *increase* the sensitivity of the limb darkening to change in filling factor and temperature gradient. So again, it would make it harder, not easier to explain an absence of limb darkening changes.

The measurements of the quiet network coverage variation over the cycle are uncertain, mainly because it is difficult to separate the quiet from active network (Foukal, Harvey & Hill 1991; Criscuoli, 2016). To estimate the effect of this uncertainty we repeated our calculations for a larger network variation of 10%. For the reason pointed out above, the CLV differences obtained for Models 2 and 3 (not shown here) remain within the measurements by Elste & Gilliam and Petro et al., while the differences obtained for Model 1 are about three times larger. So uncertainties of network coverage do not change our finding that models 2 and 3 seem preferable.

6. CONCLUSIONS

We find that the remarkable constancy of observed photospheric limb darkening does not necessarily contradict the calculated 11- year modulation of photospheric temperature gradient caused by variation in quiet magnetic network area. Such a contradiction would be suggested by the higher sensitivity of CLV to temperature gradient found in earlier calculations using 1-D and 2-D models (Petro, Foukal & Kurucz 1985). Our 3-D MHD modeling indicates that the CLV modification is within the small observed differences. So there is no fundamental contradiction between measurements of CLV variation, 11- yr. network area modulation, and of the shallower temperature gradient of flux tube atmospheres, that might reveal unexpected properties of photospheric structure.

It is unclear whether the measurements of Harder et al. (2009) provide evidence of an 11-year variation in photospheric temperature gradient. But our findings indicate that the gradient change that they report would not be inconsistent with the measured CLV variation. This is particularly true because our findings refer to the CLV and temperature gradient of the quiet sun, whereas the SIM irradiance data refer to a change in the disk averaged gradient including active network and active region faculae. Such a change in *total* facular area over a solar cycle is expected to exceed the area change of the quiet network alone by at least a factor of 2, so gradient changes of the size reported by Harder et al. would probably not be detectable in the CLV data.

Recently, variation in the photospheric temperature gradient has been reported from spectrophotometric imaging using the Solar Optical Telescope aboard HINODE by Faurobert, Balasubramanian & Ricort (2016). The authors provide a rough estimate, using a grey atmosphere approximation, suggesting that their findings may be consistent with the observed constancy of photospheric CLV. But their reported gradient difference of a few percent between solar activity maximum and minimum in 2013 and 2005 exceeds the roughly 1% network – induced gradient difference estimated in Section 2, so our 3-D MHD results suggest that it may *not* be consistent with the measured constancy of CLV.

Their results refer to changes in the inter-network, rather than in the quiet photosphere including the network, and they also exhibit lower gradient at activity maximum, so they would add to the network-induced change studied here. The sum of these effects should then be detectable in the CLV data. This discrepancy suggests that more analysis of the HINODE data is required to improve on the statistical significance of the results reported so far.

The physical mechanism that causes solar radiative variability associated with magnetic structures is now well understood (see reviews by e.g, Spruit, 1994; Foukal et al. 2006). But as explained there, photometric, helioseismic and solar diameter- based searches have found no evidence for global – scale variation in the radial temperature structure of the quiet photosphere *outside* active region or network flux tubes.

Such a finding as the HINODE results might require changes in the inter-network magnetization (e.g. Uitenbroek & Criscuoli 2014a) and of its contribution to irradiance variation. Certainly, an entirely non-magnetic origin would challenge understanding of solar heat flow because of the large thermal inertia set by the radiative upper boundary of the solar convection zone. The possibility of such a non - magnetic explanation of even the SIM results has been proposed by Harder et al. (2009). Our findings support their alternative and more plausible explanation (namely the effect of changing area of magnetic network and faculae) by showing that it cannot be ruled out by the CLV record.

Finally, the comparison carried out here might also provide a new test of 3-D MHD simulations and of radiative transfer calculations. If the variation of the network filling factor derived from observations is roughly correct, then the magnetic flux difference between simulations representing internetwork and network region is probably closer to 50 G than to 100 G. However, the relation between the filling factor in the NSO magnetograms used by Foukal, Harvey & Hill (1991) and that seen in the simulations used here is too uncertain to draw definite conclusions without further analysis.

We thank H. Uitenbroek, D. Fabbian and F. Moreno Insertis for discussion of the RH code and the MHD snapshots. One of us (PVF) thanks J. Fontenla and G. Harder for discussions of the SIM data and its interpretation, and also R. Kurucz and R. Trampedach for some initial calculations of the limb darkening sensitivity to temperature gradient. The snapshots of magneto-convection simulations were provided to us by Elena Khomenko and were calculated by F. Moreno Insertis and D. Fabbian using the computing resources of the Mare Nostrum (BSC/CNS, Spain) and DEISA/HLRS (Germany) supercomputer installations. This work was supported in part by NASA grants NNX09AP96G and NNX10AC09G to Heliophysics, Inc.

REFERENCES

Afram, N., Unruh, Y. C., Solanki, S. K., Schüssler, M., Lagg, A., Vögler, A., A&A, 526, 120 (2011)

Beck, C., Fabbian, D., Moreno-Insertis, F., Puschmann, K. G., Rezaei, R., A&A, 557, 109 (2013)

Béland, S., Harder, J., Woods, T., SPIE, 9143, 9 (2014)

Chapman, G., Ap. J. 232, 923 (1979)

Carlsson, M., Stein, R., Nordlund, A., and Scharmer, G., Ap.J. Letts 610, 137 (2004).

Criscuoli, S., Rast, M., A&A, 495, 621 (2009)

Criscuoli, S., Ap. J. 778, 27 (2013)

Criscuoli, S., Uitenbroek, H., Ap.J. , 788, 151 (2014a)

Criscuoli, S., Uitenbroek, H., A&A Letts., 562, 1 (2014b)

Criscuoli, S., Sol. Phys. , in press (2016).

Elste, G., Max Planck Institute for Astrophysics Report No 212, p. 185 (1985 )

Elste, G., and Gilliam, L., Solar Phys. 240,9 (2007).

Ermolli, I., Matthes, K., Dudok de Wit, T., Krivova, N. A., Tourpali, K., and other 10 coauthors, ACP, 13.3945E (2013)

Fabbian, D., Khomenko, E., Moreno-Insertis, F., Nordlund, Å., Ap. J., 724, 1536 (2010)

Fabbian, D., Moreno-Insertis, F., Khomenko, E., Nordlund, Å., A&A, 548, 35 (2012)

Fabbian, D.; Moreno-Insertis, F., Ap.J., 802, 96 (2015)

Faurobert, M., Balasubramanian, R. & Ricort,G., A&A in press (2016).

Fontenla, J., Avrett, E., Thuillier, G., and Harder, J., Ap. J. 639,441 (2006).

Fontenla, J. M., Harder, J., Livingston, W., Snow, M., Woods, T., JGRD, 11620108F (2011)

Fontenla, J., Stancil, P. C., Landi, E., ApJ, 809, 157 (2015)

Foukal, P., Solar Phys.120, 249 (1989)

Foukal, P., and Lean, J., Ap. J . 328,347 (1988).

Foukal, P., and Duvall, T., Ap. J. 296,739 (1985).

Foukal, P., Harvey, K., and Hill, F., Ap. J. 383, 89 (1991).

Foukal,, P., Fröhlich, C., Spruit, H., & Wigley, T. , Nature 443,161 (2006).

Garlsgaard, K., Nordlund, Å, J. Geophys. Res., 101, 13445 (1996)

Harder, J., Fontenla, J., Pilewskie, P., Richard, E., and Woods, T. GRL 36,L07801, (2009)

Hirzberger, J., Feller, A., Riethmüller, T. L., Schüssler, M., Borrero, J. M., and 12 coauthors, Ap.J. Lett., 723, 2 (2010)

Livingston, W., Wallace, L., Solar Phys 212, 227 (2003)

Livingston,W. & Milkey, R., private communication (2009)

Martínez Pillet, V., SSRv, 178, 141 (2013)

Minnaert, M., "The Photosphere". In "The Sun" G. Kuiper, ed., Univ. Chicago Press, (1953).

Neckel, H., and Labs, D., Solar Phys. 153,91,(1994).

Nordlund, Å., Stein, R. F., Asplund, M, LRSP, 6, 2 (2009).

Petro, L., Foukal, P., Rosen, W., Kurucz, R., and Pierce, A., Ap. J. 283, 426,1984.

Petro, L., Foukal, P., and Kurucz, R., Solar Phys. 98,23,(1985).

Pierce, A., and Slaughter, C., Solar Phys.51,25 (1977).

Riethmüller, T. L., Solanki, S.K., Berdyugina, S.V., Schüssler, M., Martínez Pillet, A., Feller, A. , Gandorfer, A., Hirzberger, J., A&A, 568, 13 (2014).

Rottman, G., Sol.Phys., 230, 7 (2005).

Sánchez Almeida, J., Martínez González, M., ASPC, 437, 451 (2011)

Schwarzschild,K., Nachr. Gesellsch.Wissenssch.Gottingen p.41(1906).

Schüssler, M., Vögler, A. A&A, 481L, 5 (2008).

Spruit, H., Solar Phys.50, 269 (1976)

Spruit, H., in "The Sun as A Variable Star" IAU Colloquium 142, J. Pap, ed., p. 270 (1994).

Steiner, O., Grossman-Doerth, U., Knölker, M., Schüssler, M., ApJ., 495, 468 (1998)

Steiner, O., A&A, 430, 691 (2005)

Uitenbroek, H., Criscuoli, S. Ap. J., 736, 69 (2011)

Uitenbroek, H., Ap.J., 557, 389 (2001)

Vernazza, J., Avrett,E., and Loeser, R., Ap.J. Suppl. 45, 635 (1981).

Table 1. Photospheric limb darkening observations at 445.1 nm between 1974 and 2008

| Period of observations | Reference | Activity level | Scattering correction |
|---|---|---|---|
| 3/74 – 1/75 | Pierce & Slaughter 1977 | near 1976 min. | yes |
| 9/80 - 12/82 | Petro et al. 1984 | near cycle 21 max. | yes |
| 6/86 - 6/90 | Neckel & Labs 1994 | 1986 min. - almost cycle 22 max. | yes |
| 1988, 1990 | Elste & Gilliam 2007 | rise and peak of cycle 22 | coronagraph |
| 1/02, 2/02 | Livingston & Wallace 2003 | near peak of cycle 23 | no |
| 10/07 – 6/08 | Livingston & Milkey 2007 | min. 2007-08 | no |

Table 2. Description of models

| colspan | |
|---|---|
| Maximum: 20% Network + 80% Internetwork | |
| Minimum: 15% Network + 85% Internetwork | |
| Model 1 | Internetwork: HD  Network: 100 G |
| Model 2 | Internetwork: 50 G  Network: 100 G |
| Model 3 | Internetwork: HD  Network: 50 G |

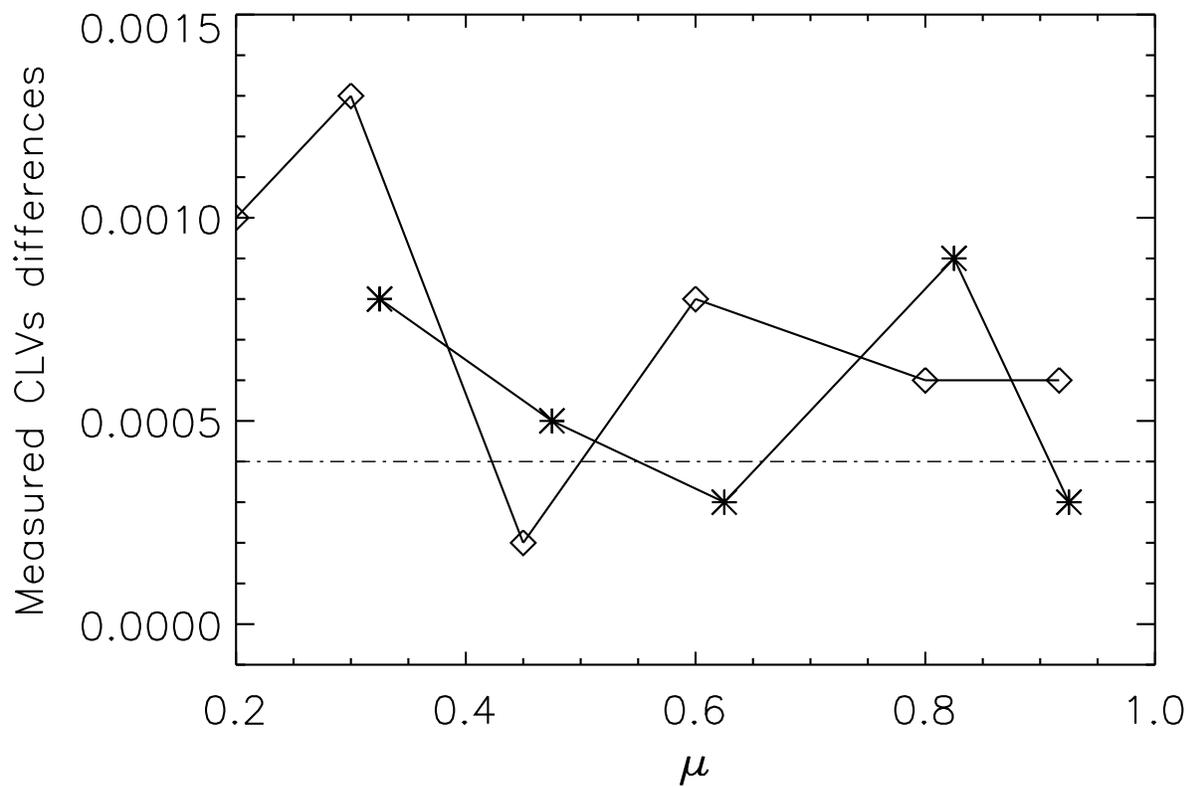

Figure 1. Differences in limb darkening at 445.1 nm versus heliocentric cosine, μ, between high and low solar activity measured by Petro et al. 1984 (stars), Elste & Gilliam 2007 (diamonds) and Neckel & Labs 1994 (dot-dashed line).

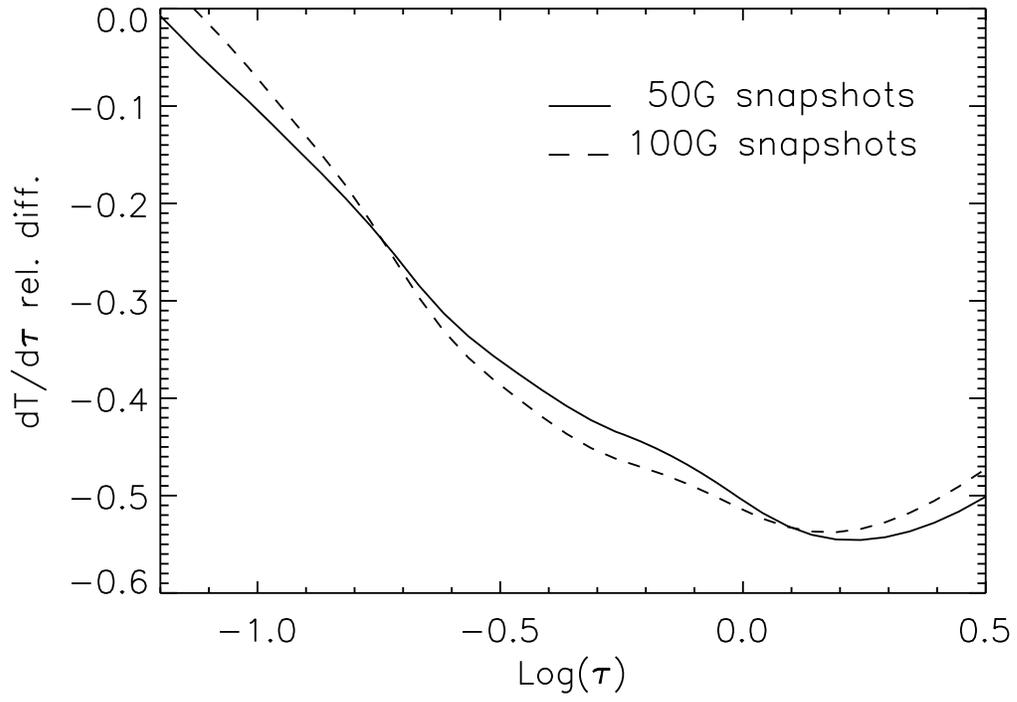

Figure 2. Relative difference between average temperature gradients of magnetic features in MHD snapshots and the average temperature gradient of the HD snapshots, versus continuum optical depth at 500 nm.

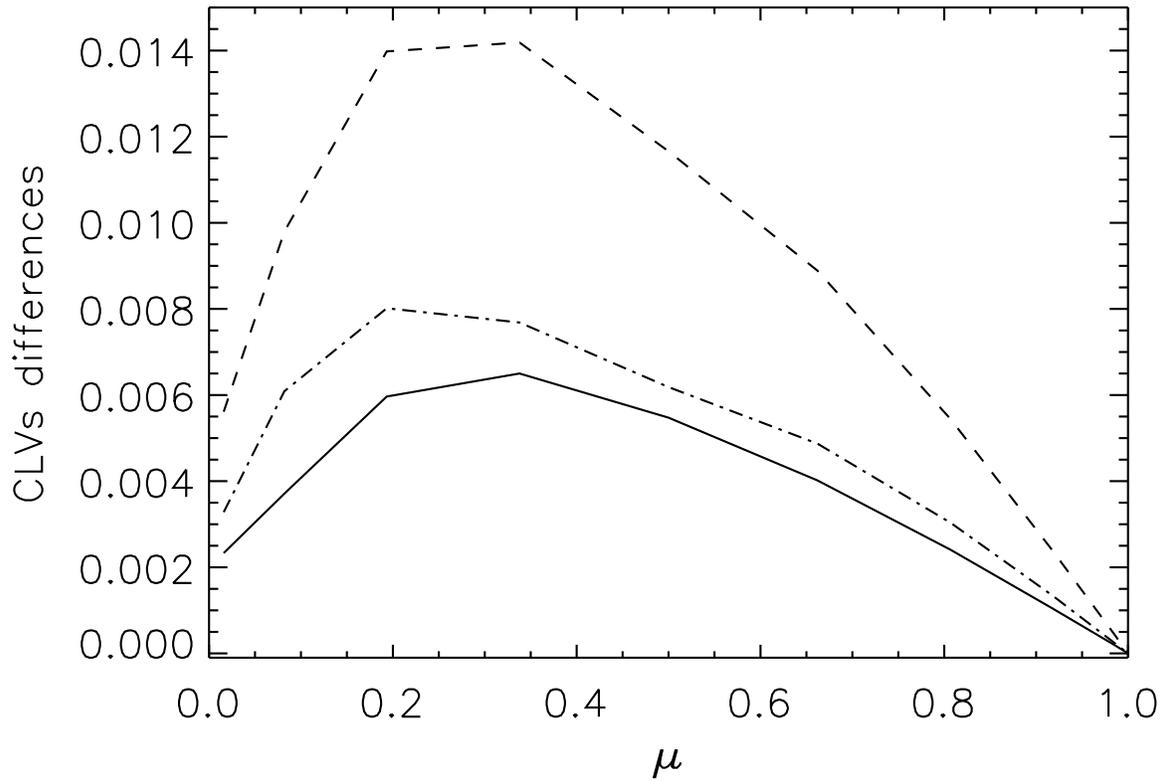

Figure 3. Calculated limb darkening differences plotted versus µ for the three sets of (M)HD simulations. Continuous: differences between 50 G and HD. Dashed: differences between 100 G and HD. Dot-dashed: differences between 50 G and 100 G.

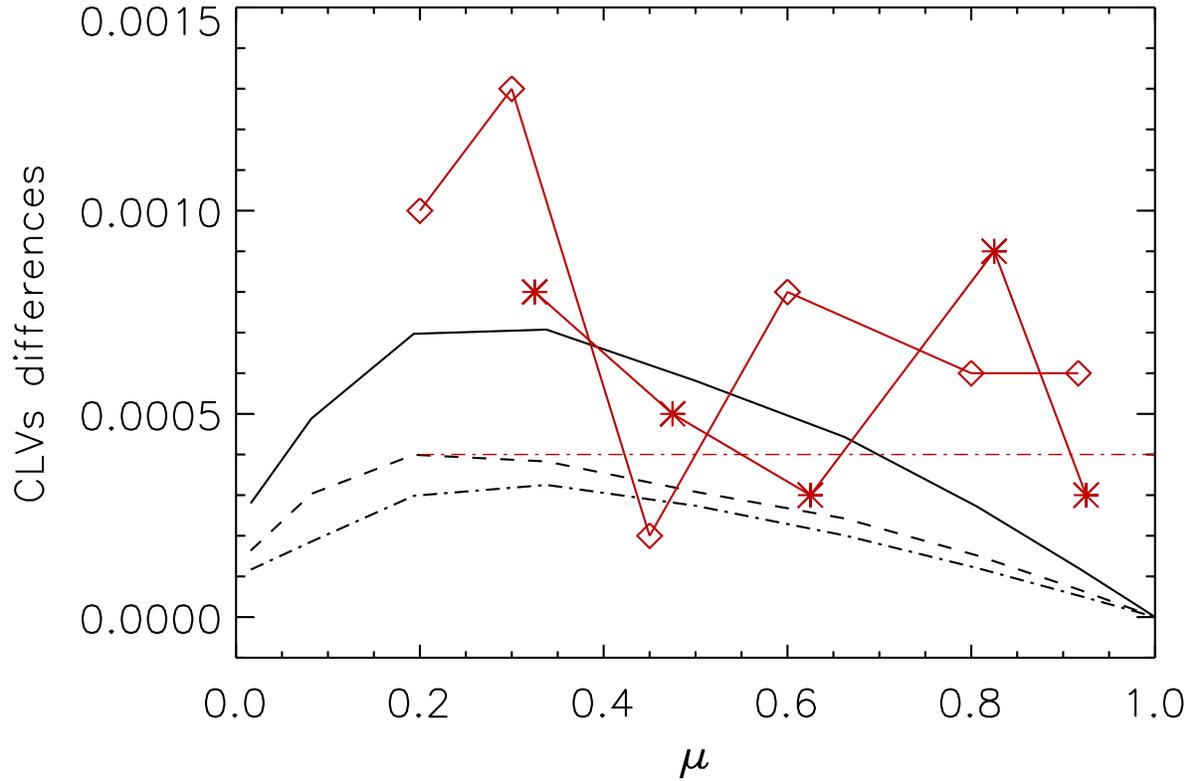

Figure 4. Calculated limb darkening differences plotted versus µ for the three network models reported in Table 2. Continuous: Model 1. Dashed: Model 2. Dot-dashed: Model 3. For comparison, results from measurements are shown in red symbols and lines as in Fig.1.